\documentclass{aa}
\usepackage{graphicx}

\begin{document}

   \title{A pair of gigantic bipolar dust jets close to the solar system}

   \author{R. Weinberger
          \and
    B. Armsdorfer}

   \offprints{R. Weinberger}

   \institute{Institut f\"ur Astrophysik, Universit\"at Innsbruck,
              Technikerstrasse 25, A-6020 Innsbruck\\
              email: Ronald.Weinberger@uibk.ac.at;
              Birgit.Armsdorfer@uibk.ac.at}

   \maketitle

   \begin{abstract}

   We present two adjacent jet candidates with a length of $\sim9\degr$
   each -- 10$\times$ longer than the largest known jets -- discovered
   by us on 60
   $\mu$m and 100 $\mu$m IRAS maps, but not observed at any other
   wavelength. They are extremely collimated (length-to-width
   ratios 20--50), curved, knotty, and end in prominent bubbles.
   Their dust temperatures are 25 $\pm$ 3 K and 30 $\pm$ 4 K,
   respectively. Both harbour faint stars, one having a high
   proper motion ($0\farcs23$ yr$^{-1}$) and being very red,
   suggesting a distance of $\sim$ 60 pc. At this distance, the
   combined mass of both jets (assuming a gas-to-dust ratio of 200) totals
   $\sim$ 1 M$_{\sun}$. We suspect that these gigantic ($\sim9$ pc length)
    jets have a common
   origin, due to the decay of a system of evolved stars. They are the
   first examples of jets radiating in the far IR and might be the
   closest non-diffuse nebulae to the solar system.

   \end{abstract}

%
%
\section{Introduction}

   Collimated jets are one of the most fascinating but poorly
   understood phenomena in astronomy. They represent ubiquitous
   features and are found in quasars, active galactic nuclei,
   young stellar objects, symbiotic systems, planetary nebulae
   and pulsars. Their
   acceleration and collimation mechanisms might be the same in all
   the classes of objects (Livio \cite{Livio}; Price et al.
   {\cite{Price}). Although an agreement on the processes that
   drive all these jets has not yet been achieved, there is
   mounting evidence that bipolar ejection is powered by
   accretion and that magnetic fields play a crucial role in
   accelerating and collimating the gas (Anderson et al. \cite{Ander}).
   The exact launching
   mechanism remains to be identified (Cabrit et al.
   \cite{Cabrit}). The main reason is the rather large
   distance of even the nearest Galactic jets which preclude
   detailed observations of regions close enough to the star
   ($<$ several AU)
   where
   the jets are generated. The distances ($\gg100$ pc) also
   impede measurements of various important features like the
   cross-section structure of jets, internal knots and other
   microstructures. In this paper we present preliminary results
   on a pair of jets which obviously is unique in several respects
   and appears to be close to the solar system.

\section{Description of the two adjacent jets}

During a survey for large dust structures around planetary nebulae
   and white dwarfs on IRAS 12 -- 100 $\mu$m
    {\em SkyView} \,
   maps we detected two new bipolar jet candidates
   at high
   latitude (b $\approx$ +67$\degr$) at 60 and 100 $\mu$m.
    The jets
   (objects A, B) are
   shown in Fig. 1.  Object A (henceforth ``A"; coined as
   ``Giant Dumbbell Nebula" by us) consists of two roundish lobes
   (A1, A2) with a diameter of $\sim1\fdg2$. They have a central
   brightness depression ($\diameter\sim0\fdg2$) in A2 and are
   connected by a wavy (amplitude $\diameter\sim0\fdg25$), narrow
   ($\diameter\sim0\fdg3$) bridge of material with a length of
   $6\fdg2$, leading to a length-to-width ratio of 21. The total
   dimension of A, whose long axis (A1--A2) has a position angle
   p.a. = $83\degr \pm 3\degr$, is $8\fdg6$. Its morphology
reminds of Herbig-Haro (HH) jets produced by young stellar
objects, except that bow-shock nebulae are frequently seen at the
ends of HH jets instead of the bubble-like structures in A. In
{\em SkyView} \, we
   found no counterpart at other wavelengths except the
   ``H${\rm \alpha}$ Composite All-Sky Survey", where an emission knot
   appears to coincide with A1 and a
   protrusion out of an extended emission structure
   appears to coincide with A2.

   In Fig. 2, A is presented at 100 $\mu$m. The white
   line follows the wiggles in the jet, showing point symmetry
   with respect to its centre (white square). The round
   terminating bubbles A2 and A1
   display a knotty appearance.

\begin{figure}
      \includegraphics[height=8.8cm]{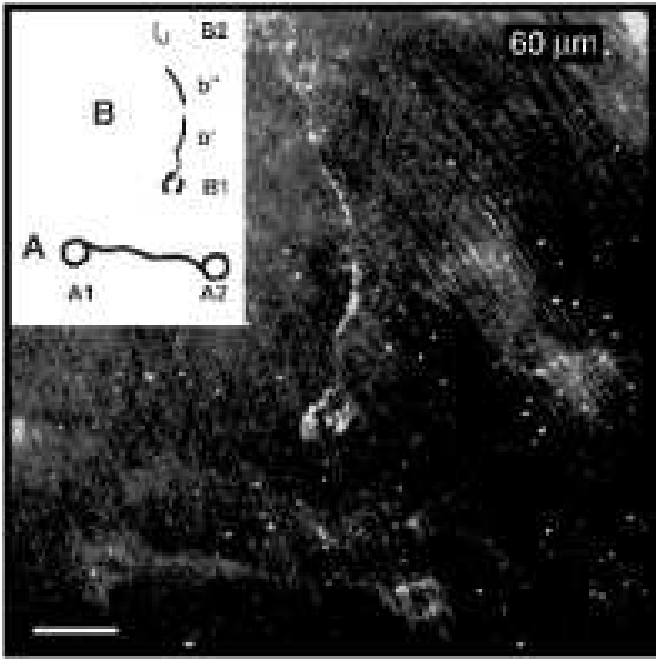}
      \caption[]{A 15$\degr\times15\degr$ field centered at
      RA = 14$^{\rm h}23^{\rm m}40^{\rm s}$, Dec =
      +40$\degr00\arcmin00\arcsec$ (J2000) in the IRAS
      60 $\mu$m passband showing both bipolar jets. The sketch at
      upper left outlines the main features of the jets which are
      referred to in the text. The bar at lower left is 2$\degr$
      long. North is at top, east to the left.
              }
         \label{FigSketchJet}
   \end{figure}

Object B (henceforth ``B"; coined as
   ``Wheat Sprout Nebula") consists of a knotty lobe
   ($\diameter\sim1\fdg2$), B1, connected to a faint thin curved
   bridge (length $\sim1\fdg4$) from B1's northern end to arc b$\arcmin$.
   The latter, of length $\sim1\fdg8$, is followed by a
   $\sim0\fdg7$ gap in emission and subsequently by arc b$\arcsec$
   which is of
   length $\sim1\fdg7$. Farther to the north there is a faint
   possible counterpart to B1, named B2; b$\arcmin$ + b$\arcsec$
   exactly match a 60$\degr$ long arc of a circle with curvature
   radius $3\fdg8$. The centre of symmetry of B is in the gap
   between b$\arcmin$ and b$\arcsec$ and there is a clear reflection
   symmetry of the parts of B with respect to this centre.
    The total dimension of B,
   whose long axis (B1--B2) has a p.a. = $4\degr \pm 1\degr$,
   is $\sim9\fdg6$. By improving the resolution of IRAS images
   from $\sim5\arcmin$  to $\sim1\arcmin$ with HIRES, the knotty
   nature of b$\arcmin$ and b$\arcsec$ becomes evident (Fig. 2).
   The five knots along b$\arcsec$ are spaced by $\sim20\arcmin$.
   The smooth southern part
   of b$\arcmin$ is of uniform width ($\sim4\arcmin$), leading to
   an extreme length-to-width ratio of 52 for b$\arcmin$ +
   b$\arcsec$. HIRES contours show the knots to be mostly double,
   displaying steeper gradients in about upstream and downstream
   directions. The knots remind of internal working surfaces in HH
   jets, caused by varying jet ejection velocities. (On Jan. 28/29 2004
   S. Temporin kindly obtained spectra with the 182 cm
   tel. of Asiago Observatory, at the position of the largest knot
    of b$\arcmin$ and of the nebula $\sim0\fdg9$ east of jet A's center.
    With regard to the [SII] 671.6+673.1 nm lines,
    we found at best traces of emission in the knot, but
  very faint emission at the nebula's position). See
    Eisl\"offel et al.
   (\cite{Eisloffel}) and Reipurth \& Bally (\cite{Reipurth}) for
   reviews of HH objects.

   In brief, A and B share the morphological characteristics with
other classes of bipolar jets, but are unique in that they are
emitting in the far IR. Their vicinity on the sky and their
comparable length, shape and brightness suggest a common origin.
The sources responsible for the ejection can hardly be young
objects, since here we deal with a very transparent Galactic
region, the Bootes Deep Field (which happens to coincide with the
centre of A) and there is no star forming region within at least
15$\degr$.

   \begin{figure}
      \includegraphics[height=7.1cm]{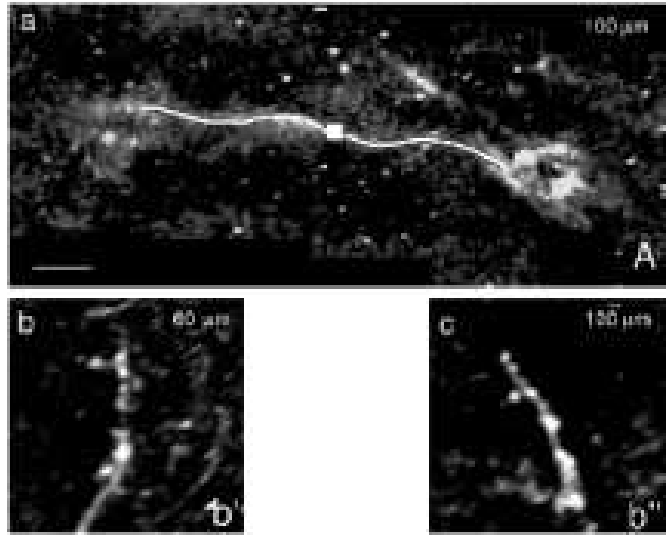}
      \caption[]{Enlargements of object A and of two arcs of
      object B. The upper panel (a) shows the southern jet at 100
      $\mu$m, using images from the IRAS server in Groningen. At
      100 $\mu$m there is a higher background contamination
      compared to 60~$\mu$m, but the curved jet structure (point
      symmetric with respect to the white square plane) is better
      visible. The square
      (RA = 14$^{\rm h}31^{\rm m}19^{\rm s}$, Dec =
      +34$\degr14\farcm2$; J2000) marks the position of the
      central star candidate; the bar is 1$\degr$ long. In the
      lower left panel (b) and lower right panel (c) parts of b$\arcmin$
      (60~$\mu$m) and b$\arcsec$ (100 $\mu$m) of object B are
      shown respectively, processed with the HIRES (IRAS) tool. Both teem with
      knots. Note e.g., at part b's northern rim,
      the slightly curved horizontal emission streak,
      whose location suggests a possible
      relation with the jet.
       Both lower panels are $2\degr \times 2\degr$.
      North is at top, east to the left.
              }
         \label{FigAandarcs}
   \end{figure}

\section{Discussion}

We searched $\sim$ 1.5 sqdeg around each jet centre (but also
within B1) for possible ejecting sources on POSS I and POSS II. We
found one stellar candidate for each jet.
 In A's very centre one faint red star
with a high proper motion (p.m.) was detected (Fig. 3). We
determined p.m. = 0$\farcs23$ yr$^{-1}$ along p.a. = 168$\degr \pm
3\degr$, and a brightness B = 20.3 $\pm$ 0.5, R = 18.0 $\pm$ 0.5.
In the 2MASS data archive we found
 J = 14.960 $\pm$ 0.039, H = 14.508 $\pm$ 0.045, and K = 14.166
$\pm$ 0.047. To estimate a distance ($D$), we note that any $D$
$>$ 100 pc leads to unreasonably high masses (and sizes) of the
jets and in addition to an uncomfortably high velocity of the star
- see below. In colour-colour diagrams, e.g. (J$-$H)--(H$-$K), or
colour-magnitude diagrams (M$_{\rm J}-$(J$-$K), M$_{\rm J}-$(R$-$J), 
etc.) (Reid \cite{Reid}) the star appears too blue for $D$ $<$100
pc. Correcting for a possible weak blue contribution
 (companion or accretion disk?) by 0.1-0.2 mag in (J$-$K) shifts
the star into the domain of late M dwarfs with M$_{\rm J}$ $\leq$
11.5 (Marley \cite{Marley}). If the corrected J is e.g. 0.3 mag
fainter (15.26) and M$_{\rm J}$ = 11.3, $D \sim$ 60 pc follows. We
will adopt this distance.

\begin{figure}
      \includegraphics[height=5.7cm]{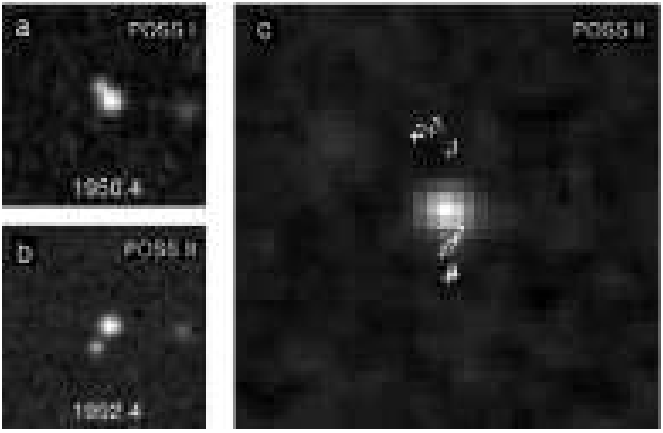}
      \caption[]{The candidate ejection sources for the two jets.
      The left panels (a, b) show the faint high-proper-motion
       star located in the centre of
      A (white square in Fig. 2), on plates of POSS\,I and
      POSS\,II, respectively. The brighter of the two stars
      probably is a background object. Both panels are
      54$\arcsec \times 54\arcsec$ wide. In the right panel (POSS\,II)
      the
      faint star in the centre of B (i.e. between arcs b$\arcmin$ and
      b$\arcsec$) is presented, together with two groups of faint
      ROSAT hri X-ray sources (nos 1-7) according to the ``ROSAT
      Complete Results Archive Sources for the HRI". However,
      these sources rather represent one source N and one S of the
      star, possibly even being only one single source. This
      panel is 36$\arcsec \times 36\arcsec$ wide. For all images
      the DSS was used. Throughout, north is top and east to the
      left.
              }
         \label{FigJetStars}
   \end{figure}

In B we searched both within the area of B1 and in the assumed jet
centre, i.e. between the arcs b$\arcmin$ and b$\arcsec$. In the
projected centre of B1, we located the eclipsing binary ROTSE1
J142344.96+374743.9. This system (GSC 2.2 magnitudes: red
$\sim12.5$, blue $\sim13.5$) however is several hundreds of pc
away. In the centre of jet B, no high proper motion star could be
detected, but one faint bluish star (B = 20.2 $\pm$ 0.5, R = 19.6
$\pm$ 0.5; RA = 14$^{\rm h}21^{\rm m}14\fs1$, Dec =
      +41$\degr54\arcmin53\arcsec$ (J2000) attracted our
      attention; it is only $\sim8\arcmin$ east of the gap's centre
      between b$\arcmin$ and b$\arcsec$ and shows a faint
      ($\sim1.6 \times 10^{-3}$ ct/s) ROSAT hri X-ray source a few
      arcsec north and south of it (perhaps even representing
      one single source coinciding with the star)
      (Fig. 3).

      For the following discussion we assume that \, i) the 60 and
      100 $\mu$m emission is thermal emission from dust, \, ii)
      dust traces gas, and \, iii) the jets are in the plane of the
      sky. We obtained dust colour temperatures $T = 25 \, \pm $ 3 K
      for A and 30 $\pm$  4 K for B. For $D$ = 60 pc, with a
      gas-to-dust ratio of 200, and using the dust mass relation
      by Hildebrand (\cite{Hildeb}), we derived total masses of
      0.99 $\pm \, 0.67$ M$_{\sun}$ for A and 0.12 $\pm$ 0.08
      M$_{\sun}$ for B. (At $D$ = 100 pc A's mass would be 2.8
      M$_{\sun}$). These low temperatures -- similar to
      interstellar cirrus temperatures --  and the lack
      of temperature gradients along the jets suggest that dust
      heating is mainly from the outside, probably from the
      interstellar radiation field. Hence, the jets might be
      cooled down fossils. The ``fragmentation" of A1, A2 and B1,
      probably resulting from instabilities,
      backs this view. Interestingly, an effective collimation
      preventing the dust from expanding must still be at work (and
      should be addressed in future studies). This mechanism -- we
      suspect magnetic fields --  however appears to allow an
      abrupt widening of the jets at their end into round lobes
      which might actually be hollow spheres, unlike the usual
      bow-shocks in HH jets. We note that at $D$ = 60 pc we are
      inside the Local Bubble (LB), where the sound speed is high
      ($\ga$ 100 km s$^{-1}$) and bow shocks may not form provided
      that the jet velocities are small enough.

      We may estimate these jet velocities. Our pair probably has
      a common origin as suggested above. This is strongly supported
      by the
      p.m. vector of A's star, $168\degr \pm 3\degr$, which must be
      of low mass
      ($\ga0.1$ M$_{\sun}$). If projected back, this
      star exactly arrives
      at the centre of B. In this centre we suspect the origin of
      the jet pair: we assume that there a triple or higher order
      stellar system had decayed. The member with the lowest mass,
      least bound to the system, has -- together with its acquired
      accretion disk -- been ejected along the plane of the sky
      and now represents the high p.m. star in the very centre of
      A. Jet A would be bent, with A's
      star at its apex (i.e. in the sense of a bow shock)
      only, if the jet has on its way
      suffered from noticeable interaction with ambient matter.
      Obviously this was not the case - as expected for a path
      within the LB.

      Provided that the ejection of both jets  started at the time
      when the stellar system broke up, then to cover the angular
  length of $8\fdg1$ between the centre of B to the centre of A
  with $0\farcs23$ yr$^{-1}$ requires 127 kyears at $D$ = 60 pc.
  The velocity of the star in the sky plane in this case amounts to 65 km
  ${\rm s}^{-1}$ (but would amount to, say, $\sim220$ km ${\rm s}^{-1}$ if the
  jets would be $D$ = 200 pc distant). Since A and B have total physical
  lengths of 9.6 pc and 9.0 pc respectively, the jet velocities
  (if constant)
  are 38 km ${\rm s}^{-1}$ and 36 km ${\rm s}^{-1}$. If these low values
  reflect the true conditions, then this explains why the jets'
  dust can survive. Further this is a hint to low mass stars
   as ejecting sources, since jet velocities are of the order of the
  escape velocities.\footnote{The referee made the following
  comment: If one star supplies the mass to the accretion disk of
  each of the two stars, one blew A and one blew B, then one of
  them (that of A) left the massive star quite early. Accretion
  disks fade quite fast -- few times the dynamical time -- after
  no mass is supplied. Therefore, it is quite possible that jet A
  was blown for a very short time only.}

  The small p.m. of the possible source of B ($>10\times$ smaller
  compared to the p.m. of the star in A) might imply a mass of
  $\geq$ 1M$_{\sun}$, in contradiction to the suggested low ejection
  velocity. Interestingly, the geometrically almost perfect
  reflection symmetry of jet B (above all, of the parts b$\arcmin$
  and b$\arcsec$) suggests an orbital motion of the ejecting
  source around a more massive object (Masciadri \& Raga \cite{Masci}).
  (The point symmetry in A can according to these authors likewise
  be taken as an indication for precession; note, that a companion
  might be necessary for the precession, but a radiation-induced
  precession might work too).

  Anyway, we could not identify this
  object -- the possible mass donor -- and speculate that it might
  be of compact nature and could perhaps even be unresolvable from
  the bluish star at the scale of the POSS. If our suggestion is
  correct, then the material of the jets must originally have
  been copiously shed via a wind by
  a massive star (this would again speak in favour
  of a quite small $D$ given b $\approx$
  +67$\degr$) and accreted by two low-mass members of the
  stellar system. The massive star
  could then, $>10^5$ years ago, have exploded as a
  supernova. Did this event contribute to the shape and hot
  gas of
  the LB? A search for a fossil wind cavity
   or supernova
  remnant led to a possible candidate: we used {\em SkyView}\,,
  selected a 50$\degr \times 50\degr$ field, centered at
  RA = 13$^{\rm h}55^{\rm m}$, Dec =
      +35$\degr$ (J2000), and used ``hist.
      equal." as brightness scaling. At 100 $\mu$m (but not
      at 60 $\mu$m)
      a huge
      ellipsoidal shell-like structure, $\sim40\degr \times
      \sim25\degr$, with its long axis at p.a. $\sim65\degr$
       becomes visible (Fig. 4). Its eastern rim is faintly present also
       in a corresponding
       nH radio map. The jet pair is located in the eastern part
       of this structure. The off-centre position might be
       due to  asymmetric mass loss of the former massive star.

       \begin{figure}
      \includegraphics[height=6.1cm]{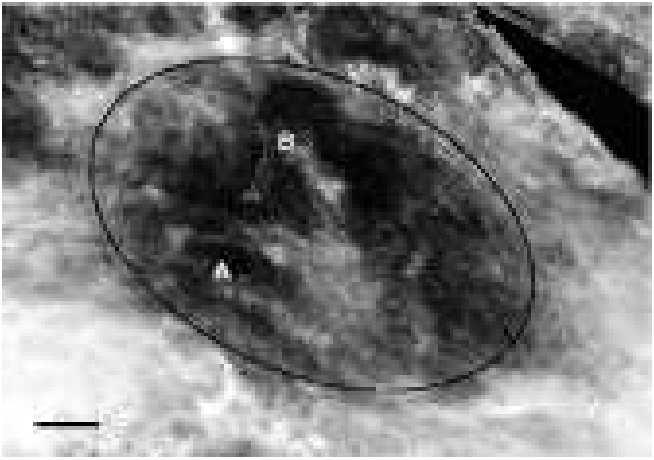}
      \caption[]{A 50$\degr\times35\degr$ field centered at
      RA = 13$^{\rm h}55^{\rm m}$, Dec =
      +35$\degr$  (J2000), i.e. l, b = 67$\degr$, +74$\degr$,
       in the IRAS
      100 $\mu$m passband showing a huge ellipsoidal shell.
      Objects A and B are marked. The bar  is 5$\degr$
      long. North is at top, east to the left.
              }
         \label{FigEllipsoidalShell}
   \end{figure}

       Alternatively, an AGB star might have
       been the mass donor. This star should however be still present as
       a bright (at least 10$^2-10^3$ L$_{\sun}$) blue object. If
       the faint bluish star close to the centre of B
       represents this young white dwarf, it must
      however be heavily obscured by the (dust) accretion disk. Indeed,
      we expect a more or less edge-on disk in both A and B, but
      the bluish nature of the candidate ejection source in B implies
      that a blue excess has to be present. Since near to this
      star two (or one)
        X-ray source(s) appear to exist (Fig. 3), such an
        excess could be explained. An other problem with
        the AGB star scenario is, that much less mass is lost
        (and consequently accreted) than
        in the former case and that a decay of a stellar system
         is less probable. A solution of this problem has to wait
         for detailed observations of the presently known candidates.

         \section{Conclusion}
         Two gigantic adjacent jet candidates, found in the far infrared,
         have been discovered. They represent the first examples of
         jets radiating in this wavelength range.
         Due to their vicinity on the sky and their
comparable length, shape, high degree of collimation and
brightness, and a proper motion of one of the candidate ejecting
stars which obviously links both jets, we suggest a common origin.
The sources responsible for the ejection can hardly be young
objects, since the jet pair is far away from any star forming
region. The physical size of the jets and the candidate ejecting
star of one of the jets with its very high proper motion suggest a
close distance, less than 100 pc, i.e. within the LB. No clear
conclusion can be drawn on the origin of the jets, but we suggest
that they stem from a decayed stellar system, where a massive star
was a mass donor and has generated accretion disks around two
low-mass stellar members. The jet pair appears to contain
cooled-down material, i.e. it is of fossil type. Other fossil jets
may exist, but will not be easily detectable because of the
ubiquitous presence of interstellar dust.

A wealth of observations will be necessary to unravel details of
these enigmatic adjacent jets. Primary goals should be to carry
out high-resolution imaging and spectroscopic observations of the
candidate ejection sources, and optical spectroscopy and CO
observations of several portions of the jets. First and foremost,
the derivation of a reliable distance should be undertaken. Apart
from understanding its very nature, this pair might be important
for future studies of the acceleration and collimation processes
in astrophysical jets due to its closeness.



\begin{acknowledgements}
      Part of this work was supported by the Austrian
      Fonds zur F\"orderung der wissenschaftlichen Forschung (FWF), project
      number P15316.
      We thank M.V. van der Sluys for fruitful discussions,
      comments and criticism. Further we are much obliged to S.
      Temporin for taking and discussing optical spectra of
      parts of jet A and jet B at the Asiago
      Observatory. We also thank B. Aryal for his help in handling
      IRAS maps and S. Kimeswenger for various comments.

\end{acknowledgements}

\end{document}